# Catalyst-Free Growth of Millimeter-Long Topological Insulator $Bi_2Se_3$ Nanoribbons and the Observation of π Berry Phase

L. Fang[1*], Y. Jia[1], D. J. Miller[1], M. L. Latimer[1,2], Z. L. Xiao[1,2*], U. Welp[1],

G. W. Crabtree[1], and W. -K. Kwok[1]

[1]Materials Science Division, Argonne National Laboratory, Argonne, Illinois 60565, USA

[2]Department of Physics, Northern Illinois University, DeKalb, Illinois 60115, USA

**ABSTRACT** We report the growth of single-crystalline $Bi_2Se_3$ nanoribbons with lengths up to several millimeters via a catalyst-free physical vapor deposition method. Scanning transmission electron microscopy analysis reveals that the nanoribbons grow along the $(11\bar{2}0)$ direction. We obtain a detailed characterization of the electronic structure of the $Bi_2Se_3$ nanoribbons from measurements of Shubnikov – de Haas (SdH) quantum oscillations. Angular dependent magneto-transport measurements reveal a dominant two-dimensional contribution originating from surface states and weak contribution from the bulk states. The catalyst-free synthesis yields high-purity nanocrystals enabling the observation of a large number of SdH oscillation periods and allowing for an accurate determination of the π –Berry phase, one of the key features of Dirac fermions in topological insulators. The long-length nanoribbons can empower the potential for fabricating multiple nanoelectronic devices on a single nanoribbon.

*Corresponding author. lfang@anl.gov
*Corresponding author. xiao@anl.gov

Topological insulators (TIs) and their potential applications have recently attracted worldwide attention.[1-3], A remarkable property of these novel material is that the charge transport of the surface state is immune to impurity scattering due to time reversal invariance. The Dirac cone of the surface states possesses linear dispersion, similar to those in graphene.[4] Of prime interest are Majorana fermions which are predicted to exist at the interface between an *s*-wave superconductor and a strong TI and can serve as the logic units of a quantum computer.[5]

Among the known topological insulating materials, $Bi_2Se_3$ has been intensely investigated due to its simple surface band structure and its large bulk band gap.[6] Angle-resolved photoemission spectroscopy (ARPES) measurements on single crystals have elucidated its two-dimensional (2D) quantum physics of the surface states.[7-9] Since bulk transport can mask surface conductance even in samples with low bulk carrier densities,[8,10,11] electrical transport experiments to delineate the surface and bulk contributions to the Fermi surface are often challenged and remain controversial. One approach to reveal the surface effect in such topological insulators is to reduce the sample size to the nanoscale, thereby significantly enhancing the surface-to-volume ratio. Indeed, Aharonov-Bohm interferences and linear magnetoresistance associated with coherent surface conduction were observed in $Bi_2Se_3$ nanoribbons.[12,13] Except for ultrathin samples obtained by exfoliation[14] with an atomic force microscope, $Bi_2Se_3$ nanoribbons have been synthesized via a vapor-liquid-solid (VLS) growth method where gold in the form of either nanoparticles[12,13,15] or a thin film[16] is used as the catalyst to promote the growth of long nanoribbons. The typical lengths of the VLS-grown nanoribbons are several micrometers, although ultra-long ribbons with lengths over 100 μm are also available.[15] However, as was revealed in VLS-grown Si nanowires, metal catalysts such as Au can unavoidably be incorporated into the synthesized nanostructures.[17-19] As a matter of fact, investigations of the Kondo effect in a TI are made possible by utilizing Au-Ni or Au-Fe double layer alloy as catalysts to grow $Bi_2Se_3$ nanoribbons, leading to the purposeful doping of Ni and Fe, respectively.[16] On the other hand, the incorporation of Au and other metal atoms can increase the bulk conductance of TI, resulting in complications in determining the surface transport state.

Shubnikov – de Haas (SdH) quantum oscillations in the magnetoresistance can provide a unique way to determine the Fermi surface and the band structures of both the bulk and surface states. The oscillations also contain information on Dirac fermions expected at the surface of a TI. The signal strength of the SdH quantum oscillations strongly depends on the quality of the sample. Up to now, SdH quantum oscillations observed in Au-catalyzed VLS-grown $Bi_2Se_3$ nanoribbons were either attributed to the bulk carriers[20] or, displayed only a small number of oscillation periods complicating the determination of crucial surface state parameters, such as the Berry phase.[13]

In this Letter, we report the *catalyst-free* synthesis of $Bi_2Se_3$ nanoribbons using a physical vapor deposition (PVD) approach. Strong SdH quantum oscillations with up to 11 cycles can be identified in the magnetoresistance of the synthesized $Bi_2Se_3$ nanoribbons in intermediate magnetic fields (≤ 9 Tesla). Analyses of the quantum oscillations reveal the dominant behavior of the two-dimensional surface states and enable us to accurately determine one of the key features of Dirac fermions on the surfaces - the π Berry phase[21,22]. Furthermore, we can grow catalyst-free $Bi_2Se_3$ nanoribbons up to a few millimeters, enabling the fabrication of multiple devices on a single nanoribbon. This long-length can benefit the integration of nanoelectronic devices, similar to those pursued by millimeter-long semiconductor nanowires.[23,24]

We used a three-zone Lindberg furnace instead of the typical single-zone furnace[12] for better temperature-gradient control in the PVD-growth of the $Bi_2Se_3$ nanoribbons. High purity polycrystalline $Bi_2Se_3$ (Aldrich, purity > 99.999%), as a source material, was placed in an $Al_2O_3$ crucible located at one end of an evacuated 16 inch-long quartz tube. The diameter near that end of the quartz tube was reduced with a natural gas torch to fix the position of the crucible. Another neck was formed on the opposite end to accommodate one piece of Si substrate (with a 200 nm $SiO_2$ layer). The entire quartz tube was pumped down to a pressure of $10^{-4}$ mTorr, then filled with high purity argon gas to a pressure of $10^{-2}$ mTorr and sealed. The quartz tube was inserted into the three-zone furnace with the $Bi_2Se_3$ source at the center of the first zone at the highest temperature for Bi$_2$Se$_3$ vaporization and the SiO$_2$/Si substrate in the center of the third zone with the lowest temperature for nanoribbon growth. The temperatures of the three zones were set to 680 C°, 620 C°, and 560 C°, respectively, resulting in a linear temperature gradient of 7.5 C°/inch across the quartz tube. The furnace dwelled at the programmed temperatures for 72 hours and was subsequently cooled down to room temperature. Large quantities of small fibrous crystals were found at the cold end, orientated randomly on the inner surface of the quartz tube. In contrast, the surface of the silicon substrate yielded only horizontally grown ribbons. Abundant ribbons of millimeter lengths were found. The typical thickness and width of the long ribbons were 1-3 micrometers and a few hundred nanometers, respectively.

The morphology and crystal structure of the synthesized Bi$_2$Se$_3$ nanoribbons were obtained using an atomic force microscope (AFM) (Park Scientific XE-HDD), a scanning electron microscope (SEM) (Hitachi S-4700), and a transmission electron microscope (TEM) (FEI Tecnai F20ST). A three-dimensional (3D) topography of a nanoribbon is shown in Fig. 1(a). Stacking features are clearly resolved at both edges, consistent with the layered crystal structure of Bi$_2$Se$_3$. In order to determine surface roughness, the tip of the cantilever was scanned over a distance of 300 nm, as depicted by the short red line in the Fig. 1(a). We found that the roughness of the surface is within 3 nm which is comparable to the c-axis lattice constant of 28.6A in Bi$_2$Se$_3$.[25] This indicates that the surface is as smooth as one single layer. The inset of Fig. 1(d) shows the TEM electron diffraction pattern indexed using a hexagonal structure. This pattern shows that the ribbon grows along the ($11\bar{2}0$) direction, which corresponds to the vertex direction of a hexagon. Interestingly, this growth direction is confirmed by a small flake grown on top of the nanoribbon, as seen in the SEM micrograph shown in Fig.1(d). A hexagon can be simply constructed with a six-fold duplication of the shape of the flake. Since the crystal structure of Bi$_2$Se$_3$ adopts a layered honeycomb-architecture, this picture clearly indicates that the ribbon grew along the vertex direction instead of the sides of the hexagon unit cell.

Nanoribbons with lengths of a few hundred micrometers were retrieved from the inner surface of the quartz tube and transferred onto ultrasonically cleaned SiO$_2$/Si substrates. Photolithographic patterning was used to define six electrical contacts along the ribbons, as shown in the inset of Fig.2(a) for standard four-probe resistivity measurements. The extra voltage pair contacts were used to probe the homogeneity of the electronic properties along the nanoribbons. The contacts were made by sputter-depositing a 50 ~ 80 nm Au layer with a 5 nm Ti adhesion layer and using a standard lift-off procedure. Two nanoribbon samples were prepared in this fashion for magnetotransport measurements. The samples were mounted onto an attocube® piezo-rotator and cooled down to cryogenic temperatures in an LHe4 variable temperature cryostat equipped with a triple-axis vector magnet (AMI) for angle dependent magnetoresistance measurements. An ac-current of 0.1 μA was provided from a lock-in amplifier (Stanford 830) and low noise current calibrator (Valhalla Scientific 2500). The magnetic field was applied in a stepwise mode and data were acquired after a dwell time of about 20 seconds. Fig. 2(a) shows the temperature dependence of the resistivity of sample #1 in zero magnetic field. We obtained a residual resistivity of ~240 μΩ cm at 2 K and residual resistance ratio (RRR) of 3.5, consistent with the semimetal behavior and low charge carrier density of this material.

As discussed above, magnetoresistance measurements have been successfully used to uncover novel surface states of a TI, for example, through the observation of Aharonov-Bohm interferences[12] and linear magnetoresistance.[13] In this work we focus on the SdH quantum oscillations which are quite pronounced in our catalyst-free grown nanoribbons to probe a key feature of a TI – the π Berry phase of the ideal Dirac fermions which are expected to exist only in high-quality samples and serve as the prerequisite to the observation of Majorana fermions.[5]

The unperturbed Dirac fermions give rise to a phase difference of π when the ground state is adiabatically subjected to a round-loop evolution.[21] In experiments, however, Berry phases less than $\pi$ have been reported in various families of TI[22] and larger than π in VLS-grown $Bi_2Se_3$ nanoribbons.[13] To understand this phenomenon, theoretical scenarios such as Zeeman coupling with the high magnetic fields and the nontrivial parabolic band dispersion of the Dirac cone[21,26] have been proposed. On the other hand, transport properties measured with superconducting electrical contacts on an exfoliated

Bi$_2$Se$_3$ crystal show a π phase shift at certain gate voltages.[27] These variations in the reported results may arise from different qualities of the sample surfaces, ambiguity due to insufficient data points, and interference from bulk transport.

The inset of Fig.2(b) shows the magnetoresistance of sample #1 obtained at various magnetic field orientations and after subtracting a polynomial background. Strong oscillatory resistance with a large number of periodic cycles can be clearly identified. SdH quantum oscillations in the magnetoresistance of a TI could originate from both surface and bulk carriers. If they arise from pure 2D surface states, their periods should depend only on the perpendicular component Bcosθ of the magnetic field, where θ is the angle between the magnetic field (B) vector and the normal direction of the nanoribbon surface. The inset of Fig.2(b) clearly shows that the oscillatory resistances follow the Bcosθ behavior of the surface states. Fast Fourier Transformation (FFT) analyses shown in the main panel of Fig.2(b) reveal only one frequency for the oscillations. Since it is unlikely that both the surface and bulk states have the same oscillation period, this indicates that the contribution from bulk carriers is negligible. As shown in Fig. 2(d), eleven SdH oscillation cycles can be clearly identified in the magnetoresistance in fields ≤ 9 T, substantially more that typically seen on Au-catalyzed VLS-grown Bi$_2$Se$_3$ nanoribbons [13,20]. The observation of a larger number of SdH quantum oscillations reveals that the electrons in our nanoribbons possess higher mobility[28], which is also confirmed by the comparison of the electron mobility 0.3926 m$^2$/Vs obtained in this work (see table I) and the reported value of 0.12 m$^2$/Vs[13]. These results indicate that our catalyst-free synthesis approach can indeed produce high-quality Bi$_2$Se$_3$ nanoribbons in which the surface states dominate the transport properties, providing an excellent platform to tackle critical issues relevant to novel phenomena in a TI.

The Landau level (LL) index n of a quantum oscillation is inversely proportional to the magnetic field, $2\pi(n+r) = S_F \hbar / eB$, where $B$ is the magnetic flux density, $S_F$ is the Fermi surface cross-section area, and the phase factor (representing the Berry phase) is $r = 0$ or $1/2$ for a regular electron gas[30] and for ideal 2D Dirac fermions[21,22], respectively. In experiments, the LL fan diagram, which plots 1/B versus n, is generally adopted to determine the Berry phase value. For ideal Dirac fermions with a linear dispersion relation, the linear extrapolation of the 1/B ~ n relationship to 1/B = 0 should intercept the n–axis at 1/2, i.e. $r$ = 1/2, corresponding to a π Berry phase. However, experimentally determined values[13,22] deviate significantly from 1/2. The open blue squares in Fig.2(c) represent the LL fan diagram of sample #1. The experimental data follow a linear relationship with an n-axis intercept at n = 1/2, as expected for a π Berry phase. The precision of this analysis can be further attested by fitting the entire oscillatory resistance with that predicted by the Lifshitz-Kosevich (LK) theory for a two-dimensional generalized Fermi liquid with the addition of an extra Berry phase.[4,13,28-31] This theory gives the following formula: $\Delta R \propto R_T R_D \cos[2\pi(F/B + \frac{1}{2} + r)]$ for the oscillatory resistance, where $R_T = \alpha T m_c^* / B \sinh[\alpha T m_c^* / B]$ is the thermal factor, $R_D = \exp[-\alpha T_D m_c^* / B]$ is the Dingle damping factor with α = 14.69 T/K, and $T_D = \hbar / 2\pi k_B \tau$ is the Dingle temperature, and the effective cyclotron mass (in units of the free-electron mass $m_e$) is given as the change in Fermi surface cross section $S_F$ with energy, $m_c^* = \hbar^2 / 2\pi m_e \, \partial S_F / \partial \varepsilon$. The excellent match between the experimental data (open circles) and the theoretical curve (blue line) with a phase factor of $r$ = 1/2 as shown in Fig. 2(d), clearly indicates the existence of a π Berry phase.

As discussed above, it is very challenging to experimentally determine the Berry phase: the sample must be of excellent quality so that pronounced quantum oscillations can be observed at intermediate values of the magnetic field to avoid the interference from spin-splitting at high magnetic fields.[22] More importantly, the bulk transport must be suppressed so that the surface states dominate the observed quantum oscillations. In experiments, a π Berry phase has only been found in an exfoliated Bi$_2$Se$_3$ crystal at certain gate voltages.[27] In Au-catalyzed VLS-grown Bi$_2$Se$_3$ nanoribbons, the fit of oscillatory magnetoresistance to LK theory yields a Berry phase of (1.62±0.02)π.[13] Below, we will show that even a very weak contribution from bulk transport can affect the determination of the Berry phase.

Figure 3(a) presents the angle dependent oscillatory resistances for sample #2. In order to visualize the contribution of the surface states, we again used the perpendicular field component, Bcosθ, as the variable. The left dashed line in the figure shows that the minima in the magnetoresistance curves obtained at large angles (θ ≥15°) do not align with those at small-angles, indicating the possible contribution from the 3D bulk Fermi surface. In fact, we were able to resolve two different types of wiggles in the oscillatory resistance, as indicated with arrows and stars in Fig.3(a). The FFT analysis in Fig. 3(b) reveals a sharp major peak ($\alpha$1) and a minor peak ($\alpha$2) at lower frequency. Both peaks move to higher frequency with increasing angle, while their amplitude decreases. The angle dependent frequencies of $\alpha$1 and $\alpha$2 are plotted in Figs.3(c) and (d) respectively. The 1/cosθ dependence of the $\alpha$1 peak indicates that it originates from a 2D band structure. As a comparison, we calculated the angle dependent frequency based on the 3D effective mass anisotropy[11] of the bulk conduction band (CB) of $Bi_2Se_3$ and plotted the curve in the same figure. This curve deviates significantly from the $\alpha$1 curve at θ ≥30°. On the other hand, the angle dependent frequency of the $\alpha$2 peak can be described by models based on either the 2D surface states or 3D effective mass anisotropy, within the experimental uncertainty. Since a 2D surface cannot have two different frequencies and the $\alpha$1 peak has already been determined to originate from the surface band, we attribute the $\alpha$2 peak to the bulk conduction band of $Bi_2Se_3$. This identification is in fact consistent with the derived band structure for this sample. (see section B in supporting information).

The LL fan diagram of sample #2 is given in Fig.2(c) as red open circles. Due to the weak bulk contribution as revealed by the FFT analysis, the data points in the fan diagram show relatively higher fluctuations about a straight line compared with that of sample #1. The best linear fit to the data points results in an intercept at n = 0.45, i.e. $r$ = 0.45, corresponding to a Berry phase of 0.9π. However, this does not imply that the Dirac fermions in this nanoribbon are non-ideal. In fact, as demonstrated by Fig. 3(e), the curve calculated from the LK theory with a phase factor $r$ = 1/2 and the determined $\alpha$1-frequency matches the prime periodicity of the oscillatory resistance quite well: the locations of the first, third, fourth and sixth resistance minima (counted from the left) of both the theoretical and the experimental curves are nearly at the same location while slight deviations in the second, fifth and seventh minima can be induced by the superposition of the low-frequency bulk component. That is, ideal Dirac fermions most likely still exist in this nanoribbon. These results indicate that precise determination of the Berry phase based on the LL fan diagram can be perturbed by a weak bulk contribution. Such an uncertainty can be further compounded by a limited number of oscillation cycles, and hence highlights the desirability of high-quality $Bi_2Se_3$ nanoribbons.

Table I summarizes the derived physical parameters for the surface states of both samples #1 and #2 (see supplemental information). Interestingly, the values of the quantum oscillation frequency F, Fermi vector $k_F$, sheet electron density $n_{2D}$, and effective mass $m_c^*$ for sample #2 are comparable to those of the Au-catalyzed VLS grown nanoribbons with F = 86,[20] $n_{2D}$ = 1.3 x$10^{12}$/cm$^2$, $m_c^*$ = 0.12$m_e$, and $k_F$ = 0.41 nm$^{-1}$ [Ref.13] while the mobility is a factor of 2.5 larger than the reported value of 0.12 m$^2$/Vs.[13] On the other hand, sample #1 has more pronounced surface effects with two times larger F, $n_{2D}$, scattering time $\tau$ and mean free path $l$. The larger $\tau$ and $l$ values denote less scattering on the surface, enhancing the SdH quantum oscillations. The different sheet electron densities in these two samples could originate from different degrees of air exposure, which has been found to result in additional n-type doping. In fact, sample #1 was stored longer in a nitrogen box, although both samples experienced the same open-air electrical contact procedure. However, an air exposure of 1-2 hours was found to smear the SdH oscillations significantly.[20] This indicates that there is a delicate balance between doping-induced increase of the sheet electron density to enhance the surface state signals and oxidation, which can deteriorate the sample.

In a bulk TI, the conductance of bulk carriers could easily overshadow that of the surface channel.[32-34] Thus, it is important to achieve low bulk carrier density. The detectable bulk component ($\alpha$2), though weakly exhibited in the SdH oscillations of sample #2 enabled us to estimate the bulk carrier density in this sample. Using the free electron relation[35] $k_F = (3\pi^2 n)^{1/3}$, we obtained $n_{3D}$ = 3 × $10^{17}$ /cm$^3$. This value is more than a factor of 10 smaller than that of the Au-catalyzed VLS-grown nanoribbons, markedly demonstrating the advantages of our catalyst-free synthesis approach in suppressing the bulk transport.

The millimeter-long $Bi_2Se_3$ nanoribbons should provide enough length to fabricate multiple quantum-computing devices on a single nanoribbon. The homogeneity of a long ribbon thus becomes another critical factor to affect the performance. We find the SdH oscillatory resistance among different voltage pairs to be virtually identical, indicating good electronic homogeneity of our catalyst-free $Bi_2Se_3$ nanoribbons.

In conclusion, we successfully synthesized nanoribbons of the topological insulator $Bi_2Se_3$ with millimeter-scale lengths using a *catalyst-free* physical vapor deposition method. These ultra-long nanoribbons grow along the $(11\bar{2}0)$ direction and have typical widths and thicknesses of few micrometers and hundreds of nanometers, respectively. We observed strong Shubnikov – de Hass (SdH) quantum oscillations in the magnetoresistance. Magnetic field orientation dependent measurements and the observation of a vanishing but still resolvable bulk component enabled us to identify the topological surface states as the dominant contributor to the SdH quantum oscillations. The fan diagram of the Landau levels and the fitting of the oscillatory resistance with the Lifshitz-Kosevich theory yield a Berry phase with a phase quantity π, revealing the existence of ideal Dirac fermions in the topological insulator $Bi_2Se_3$. Our ultralong nanoribbons have uniform electronic properties, enabling potential applications in the fabrication of multiple nanoelectronic devices on a single nanoribbon.

**Acknowledgement.** This work was supported by the Department of Energy, Office of Basic Energy Sciences, under Contract No. DE-AC02-06CH11357 (DJM, MLL, ZLX, UW, GWC, WKK). Crystal synthesis was supported by the Center for Emergent Superconductivity, an Energy Frontier Research Center funded by the US Department of Energy, Office of Science, Office of Basic Energy Sciences (LF, YJ). The SEM and TEM analyses were performed at Argonne's Electron Microscopy Center (EMC), which is funded by the US Department of Energy under contract DE-AC02-06CH11357.

**Supporting information available.** Cyclotron mass and band structure of sample #2.  This material is available free of charge via the internet at http://pubs.acs.org.

**Figure Captions**

**Fig.1**. Morphological and structural characterizations of the PVD-grown nanoribbons. (a) AFM micrography of a typical $Bi_2Se_3$ nanoribbon. (b) and (c) SEM top- and side-views of a nanoribbon, revealing its width of ~ 400 nm and thickness of ~ 150 nm. (d) SEM micrograph of a nanoribbon with a small flake grown on the (*0001*) plane. The hexagon drawn around the flake visualizes the growth direction of the nanoribbon. The TEM diffraction pattern is shown in the inset highlighted by a yellow square. (e) A photograph of a nanoribbon longer than 1 mm grown horizontally on the $SiO_2$/Si substrate. Its width and thickness are 1.4 um and 200 nm, respectively.

**Fig.2**. Determination of the Berry phase. (a) Temperature dependence of the resistivity for sample #1. The inset is a SEM micrograph of a nanoribbon with six photolithographically patterned electrical contacts. (b) Fast Fourier Transformation (FFT) of the oscillatory resistance ($\Delta R$) for sample #1 in perpendicular fields and at 2 Kelvin. The inset shows $\Delta R$-curves obtained at various magnetic field orientations. The normal direction of the nanoribbon surface is defined as $\theta = 0°$. (c) The fan diagram of Landau-levels for both samples #1 and #2. (d) Comparison of the experimental data points of sample #1 with the calculated curve based on the LK theory with a phase factor of $r = 0.5$.

**Fig.3**. Analysis of the SdH quantum oscillations in sample #2. (a) Oscillatory resistance ($\Delta R$) curves obtained at different angles plotted as a function of $1/B\cos\theta$. The stars and arrows highlight two different types of waveforms. (b) FFT data of the oscillatory resistance showing two distinct frequency peaks, $\alpha 1$ and $\alpha 2$. (c) and (d) Angle dependence of the two frequencies. The symbols are experimental data and curves are calculated. See text for more details. (e) Comparison of the experimental data and the calculated curve based on the LK theory with a phase factor of $r = 0.5$.

# FIGURE 1

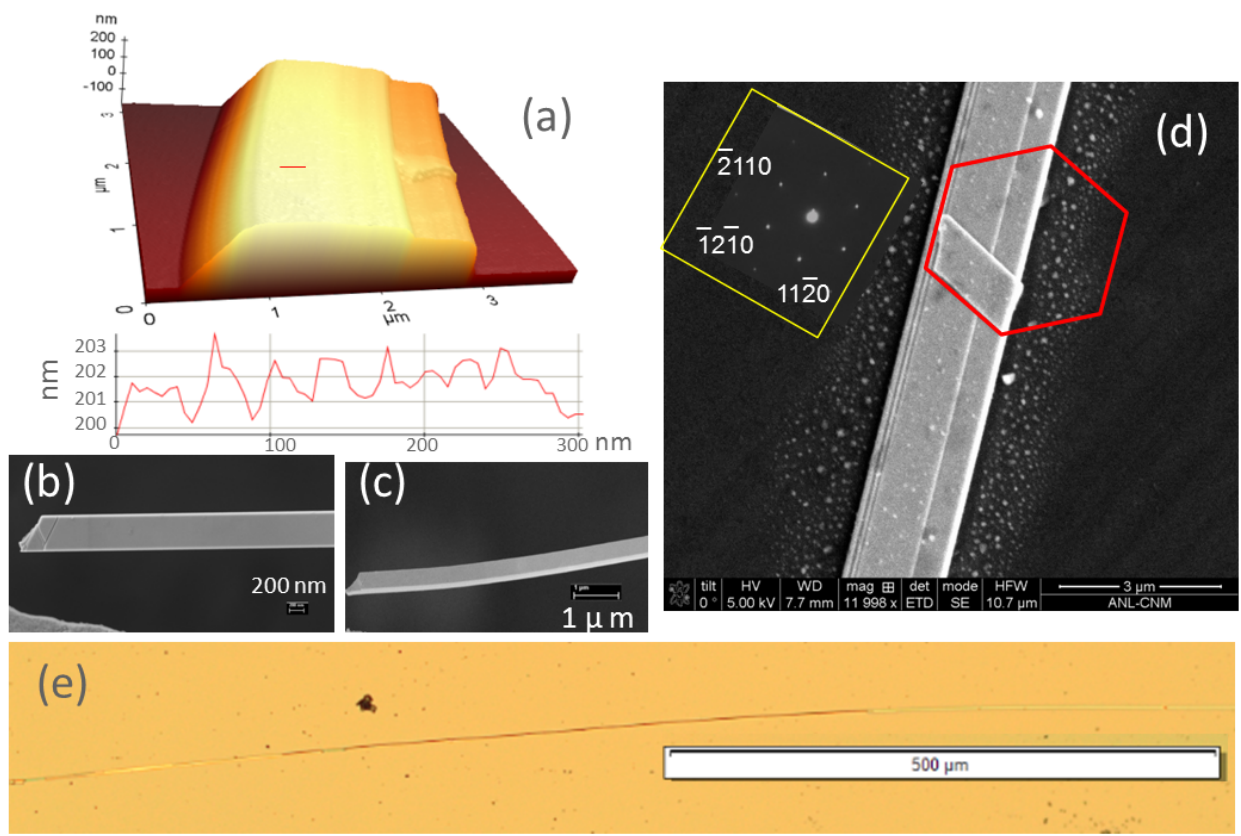

**FIGURE 2**

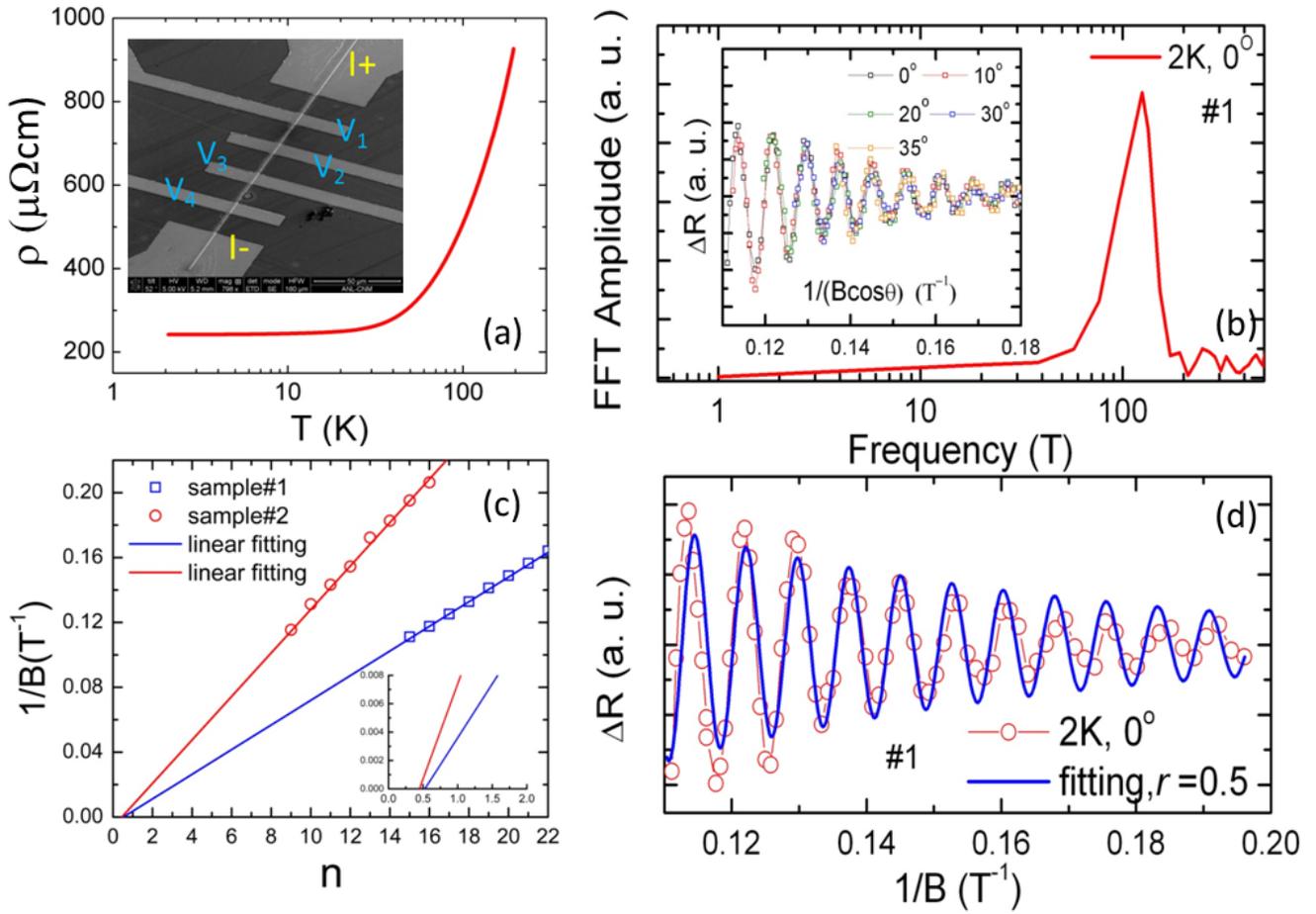

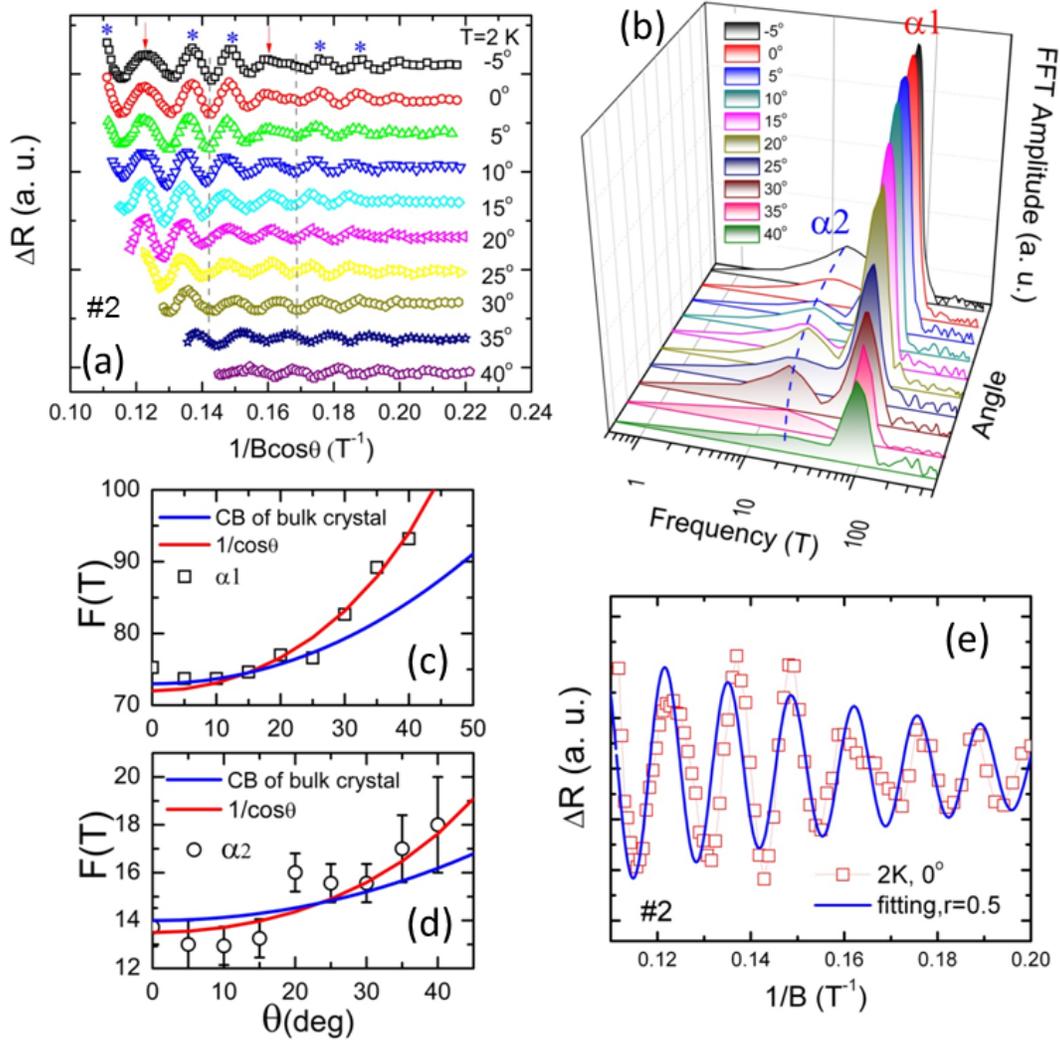

**Table I.** Physical parameters of the surface states of the measured $Bi_2Se_3$ nanoribbons

| Sample | F (T) | $k_F$ (nm$^{-1}$) | $n_{2D}$ ($10^{12}$cm$^{-2}$) | $m_c^*$ ($m_e$) | $v_F$ ($10^5$ m/s) | $E_F$ (meV) | $\tau$ ($10^{-13}$s) | $l$ (nm) | $\mu$ (m$^2$/Vs) |
|---|---|---|---|---|---|---|---|---|---|
| #1 | 131 | 0.63 | 3.16 | 0.18±0.1 | 3.96 | 176 | 4 | 161.7 | 0.3926 |
| #2 | 74 | 0.47 | 1.786 | 0.14±0.1 | 3.8 | 128 | 2.4 | 92.6 | 0.302 |

# Supporing information

### A. Determination of the cyclotron mass

The cyclotron effective mass (in units of $m_e$) is $m_c^* = \frac{\hbar^2}{2\pi m_e}\frac{\partial S_F}{\partial E}$, where $E = a\hbar k$ ($a$ is the initial velocity) is the energy at wave vector $k$ and $S_F = \pi k^2 = \pi\left(\frac{E}{a\hbar}\right)^2$ is the cross-sectional area of $k$-space orbits at wave vetor $k$. For linear dispersion, the cyclotron mass is then given as $m_c^* = \frac{\hbar k}{m_e a}$, which means, the effective mass is zero at the Dirac point $k = 0$ ($E = 0$), however, increases at higher energy/velocity.

The cyclotron mass ($m_c^*$) at the Fermi velocity of the surface ($\alpha 1$) band can be extracted from the temperature dependence of the oscillatory resistances at various fixed magnetic fields. The LK theory predicts a thermal factor, $R_T = \alpha T m_c^* / B \sinh[\alpha T m_c^* / B]$. A comparison between the theory and experiment, as shown in Fig. S1, yields $m_c^* \approx 0.14 m_e$. The inset of Fig. S1 presents the LK fitting for data obtained at another field, yielding a value of $\sim 0.13 m_e$. These values are consistent with those reported in the literature.[1]

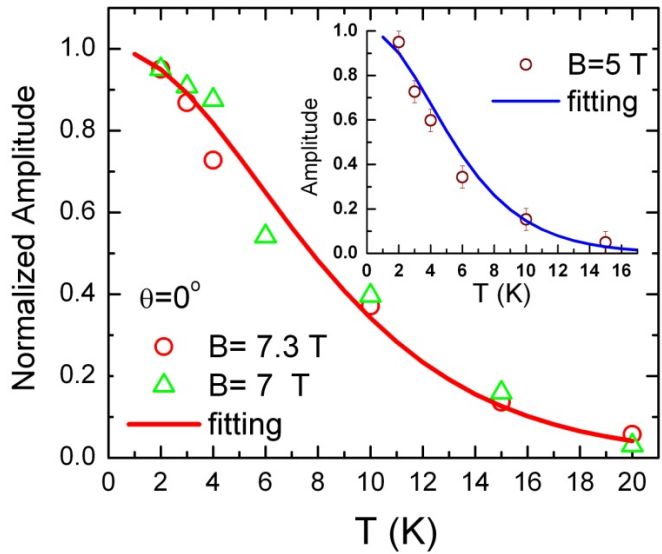

**Fig. S1**.

Temperature dependence of the thermal factors. The symbols are experimental data while the curves are fits with LK theory. The inset is the same plot for data at another magnetic field.

**B. Construction of the band structure for sample #2**

Using the Onsager relation for a circular Fermi surface cross sections, $F=(\hbar/2\pi e)\pi k_F^2$, where $F$ is the frequency of the oscillatory resistance and $\hbar$ is Planck's constant divided by a factor of $2\pi$, we obtain the Fermi wave vector $k_F$=0.048/$\overset{o}{A}$ and 0.021/$\overset{o}{A}$, for the $\alpha$1 and $\alpha$2 peaks, respectively. The LK fitting yields cyclotron mass at $k_F$ $m_c^*$ =0.14 $m_e$ for the surface band (see section A). The Fermi velocity, derived from the expression of cyclotron effective mass, is $v_F = \dfrac{(2e\hbar F)^{1/2}}{m_e m_c^*} = 3.8 \times 10^5 \, m/s$, corresponding to a Fermi energy of $E_F = v_F \hbar k_F$ = 128 meV.

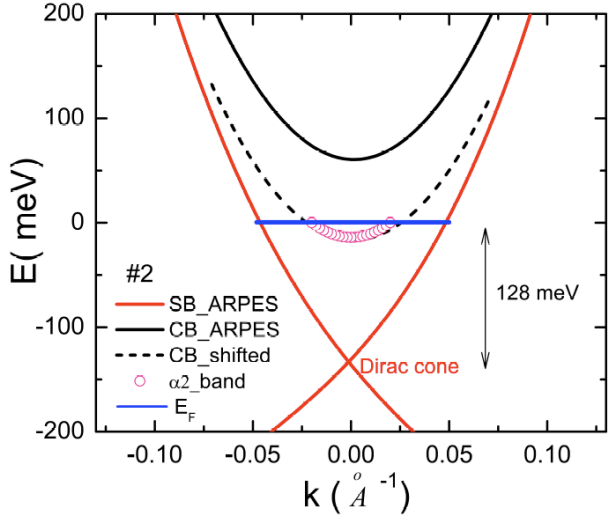

Fig. S2 Caluculated band structure for sample #2.

The Fermi surface of this sample is thus located at about 128 meV above the Dirac point. As for the $\alpha$2 peak, we associate it with the bottom of the conduction band. Assuming weak correlation effects in the bulk of $Bi_2Se_3$, the quasiparticles behave approximately as a free electron gas. The dispersion of the conduction band is then approximately quadratic $\hbar^2 k_F^2 / 2m_e^*$. The effective mass of the carrier in the conduction band $m_e^*$ has been determined to be 0.125 $m_e$.[2-3] This yields an energy scale of ~ 14 meV. We thus construct the band associated with the $\alpha$2-peak as an electron pocket with a wave vector 0.021/$\overset{o}{A}$ and the bottom of the pocket located 14 meV below $E_F$.

Based on band structure calculations and ARPES measurements,[4-6] the bottom of the conductance band of $Bi_2Se_3$ is about 205 meV above the energy level of the Dirac cone. However, the band associated with the α2-peak in the present study is located at 128 meV above the Dirac point. This difference could stem from the so-called band-bending scenario. Transport studies on the bulk $Bi_2Se_3$ crystal found that the surface becomes *n*-doped once it is exposed to an ambient environment. Consequently, a downward band bending of about 75 meV from the surface towards the bulk was inferred.[6] For a better comparison, we extracted the band dispersion of the bulk $Bi_2Se_3$ crystal from the literature[4,5] and plotted them in Fig. S2. The solid blue line, 128 meV above the Dirac point, represents the $E_F$ of our sample. The constructed electron pocket (the α2-peak associated band) is also plotted in the figure. We find that if the conduction band is lowered by 77 meV, it fits the electron pocket remarkably well. The reduced energy is virtually identical to the reported band bending 77 meV.[6] Thus, we conclude the α2-peak originates from the bulk conductance band of the $Bi_2Se_3$ nanoribbon.